\documentstyle[12pt]{article}
\begin{document}
\textheight 220mm
\textwidth 140mm
\topmargin -2mm
\renewcommand{\thefootnote}{\ast}
\newcommand{\be}{\begin{eqnarray}}
\newcommand{\beq}{\begin{equation}}
\newcommand{\ba}{\begin{array}}
\newcommand{\ee}{\end{eqnarray}}
\newcommand{\eeq}{\end{equation}}
\newcommand{\ea}{\end{array}}
\newcommand{\zt}{\zeta}
\newcommand{\ve}{\varepsilon}
\newcommand{\al}{\alpha}
\newcommand{\gm}{\gamma}
\newcommand{\Gm}{\Gamma}
\newcommand{\om}{\omega}
\newcommand{\et}{\eta}
\newcommand{\bt}{\beta}
\newcommand{\dt}{\delta}
\newcommand{\Dt}{\Delta}
\newcommand{\La}{\Lambda}
\newcommand{\la}{\lambda}
\newcommand{\vp}{\varphi}
\newcommand{\nn}{\nonumber}
\newcommand{\nid}{\noindent}
\renewcommand{\baselinestretch}{1.4}
\newcommand{\lmx}[1]{\begin{displaymath} {#1}=
                    \left(\begin{array}{rrr}}
\newcommand{\rmx}{\end{array} \right) \end{displaymath}}

\begin{titlepage}
\begin{center}
  {\Large \bf The stability of a cubic fixed point in three 
   dimensions from the renormalization group}
\end{center}
\vspace{0.2cm}
\begin{center}
{\bf K. B. Varnashev\footnote{E-mail address: feop@eltech.ru}}
\end{center}
\begin{center}
Department of Physical Electronics, Saint Petersburg 
Electrotechnical University, \\
Professor Popov Street  5, St.Petersburg, 197376, Russia
\end{center}

\begin{center}
{\bf  Abstract}
\end{center}

The global structure of the renormalization-group flows of a model with
isotropic and cubic interactions is studied using the massive field theory
directly in three dimensions. The four-loop expansions of the $\bt$-functions
are calculated for arbitrary $N$. The critical dimensionality
$N_c=2.89 \pm 0.02$ and the stability matrix eigenvalues estimates obtained 
on the basis of the generalized Pad$\acute{\rm e}$-Borel-Leroy resummation
technique are shown to be in a good agreement with those found recently by
exploiting the five-loop $\ve$-expansions.

\vspace{0.5cm}
\centerline{(1 December 1999)}

\vspace{1.5cm}
\nid
Published in: {\it J. Phys. A: Math. and Gen. {\bf 33} (2000) 3121-3135.}

\vspace{5.5cm}
\rightline{{\sl Typeset using} \LaTeX}
\end{titlepage}

\section{Introduction}
\label{sec:1}

The study of the critical properties of magnetic phase transitions
in three dimensional (3D) cubic crystal is a problem attracting
theoretical efforts over more than 25 years.
By using the lower-order renormalization-group (RG) approach, Wilson
and Fisher \cite{WF72}, Aharony \cite{Ah73}, and Ketley and Wallace
\cite{KW73,W73} showed that in the
critical region the fluctuation instability of continuous phase transitions
may be observed, and that it may lead to the isotropization of the system
with a cubic anisotropy. This fact gave rise to the question of what regime 
of the critical behaviour is actually realized in 3D cubic crystal with $N=3$.
It was soon understood that the calculation of the critical dimensionality
$N_c$ of the order parameter is the crucial point in studying critical 
phenomena in a cubic crystal.
Indeed, the critical value $N_c$ separates two different regimes of critical
behaviour of the system. For $N > N_c$ the cubic rather than the isotropic
fixed point is stable in 3D. At $N = N_c$ the points interchange their
stability so that for $N < N_c$ the stable fixed point is the isotropic one.
However, attempts to evaluate the critical dimensionality
resulted in dramatically different estimates.

In fact, the one-loop RG analysis of the stability matrix eigenvalues of
the cubic and isotropic fixed points as well as some symmetry arguments
(see section 2) predicts that $N_c$ should lie between 2 and 4.
Many years ago the three-loop expansion for $N_c$ as a power series in
$\ve$ was obtained \cite{KW73}.
Summation of that short series at $\ve = 1$ (D = 3) by means of the
Pad$\acute{\rm e}$ approximant [1/1] yielded the value $N_c = 3.128$
\cite{Aha}, while making use of the Pad$\acute{\rm e}$-Borel resummation
method results in the estimate $N_c=3$. In contrast to this, the value
$N_c = 2.3$ has been found on the basis of the variational modification
of the Wilson recursion relation method in \cite{YH77}.
Later, however, Newman and Riedel showed by decoupling the infinite
system of the recursion relations for the scaling fields and
then solving them that for $D = 3$ $N_c \sim 3.4$ \cite{NR82}.
At the same time, the classical technique of high-temperature expansions,
under some circumstances, allowed one to establish that for $N = 3$ 
the isotropic
critical asymptotics in a cubic crystal is unstable \cite{FVdC81},
thus implying $N_c < 3$.
Ten years ago the analysis of the critical behaviour of the
(mn)-component field model, which has a good number of interesting
applications to the phase transitions in real substances, has been carried
out within the three-loop RG approach in three-dimensions. The calculation
of the stability matrix eigenvalues for the cubic model ($m = 1$,
$n = 3$) provided the stability of the cubic fixed point in 3D, and the
critical dimensionality turned out to be equal to 2.91 \cite{Sp89}.
In agreement with this, the estimate $N_c = 2.9$ was given in \cite{MSSh89}.
More recently, Kleinert and Schulte-Frohlinde calculated the RG functions
for the cubic model in $(4-\ve)$ dimensions up to five-loop order \cite{KS95}.
Summation of the critical dimensionality expansion with the help of the
Pade approximant [2/2] gave the estimate $N_c = 2.958$ \cite{remark}.
The cubic fixed point eigenvalues found by means of a simple resummation
algorithm of the Borel type, accounting for the large-order behaviour of the
$\bt$-functions when the anisotropy parameter is very small
\cite{PRd95}, indicated that the cubic point is stable in 3D \cite{KTSf-b}.
Finally, in the recent work Ref. \cite{CH98} by using finite size scaling
techniques and the high precision Monte Carlo (MC) simulation it has been
suggested that $N_c$ coincides with three exactly. Such strong scattering
in the estimates of $N_c$ motivated us to study this problem
with particular care. Calculation of the critical dimensionality
as well as the eigenvalue exponents for the cubic and isotropic fixed
points by exploiting the higher-order RG approach in three dimensions and
generalized Pad$\acute{\rm e}$-Borel-Leroy (PBL) resummation technique is the
main goal of the paper. As will be shown, our estimates for $N_c$ and
eigenvalues are in excellent agreement with recent results \cite{KTSf-b}
obtained on the basis of the five-loop $\ve$-expansions.

The layout of the paper is as follows. In the next section the model
Hamiltonian is introduced and the massive field-theoretical RG procedure
in fixed dimensions is formulated. The perturbative expansions
for $\bt$-functions for generic $N$ are then deduced up to the four-loop 
order. In section 3 the structure of the RG flows of the model are 
investigated and the fixed point locations are calculated using the 
generalized PBL
resummation method. The eigenvalue exponents of the most intriguing
$O(N)$-symmetric and cubic fixed points are evaluated for the physically
significant case $N=3$ and the stability problem is solved.
The numerical estimate of the
critical dimensionality $N_c$, at which the topology of the flow diagram
changes, is obtained by resumming both the four-loop RG expansions for the
$\bt$-functions in 3D and the five-loop $\ve$-expansion for $N_c$ at $\ve=1$.
In the conclusion the results of the investigation are discussed, along with
the predictions and numerical estimates obtained earlier on the basis of
the same or other theoretical approaches.

\section{The model, RG procedure and $\bt$-functions}
\label{sec:2}

We start from the fluctuation Hamiltonian
\be
H =
\int d^d x \Bigl[{1 \over 2}( m_0^2 \vp_i^2
 + \partial_\mu \vp_i \partial_\mu \vp_i)+
{1 \over 4!} \Bigl(u_0 G_{\>ijkl}^{\>1}+
v_0 G_{\>ijkl}^{\>2}\Bigr)
\vp_i \vp_j \vp_k \vp_l\Bigr] \ , \label{eq:1}
\ee
where $\vp_i$, $i = 1, \ldots, N$, is the real vector order parameter 
field in fixed $d$ and $m^2_0$ is the linear measure of the temperature, 
$u_0$ and $v_0$ denote the "bare" coupling constants. The symmetrized 
tensors associated with isotropic and cubic interactions are
\be
 G_{\>ijkl}^{\>1} = \frac{1}{3}\Bigl(
\delta_{ij}\delta_{kl}+
\delta_{ik}\delta_{jl}+
\delta_{il}\delta_{kj}\Bigr) \ , \qquad
 G_{\>ijkl}^{\>2} =
\delta_{ij}
\delta_{ik}
\delta_{il} 
\label{eq:2}
\ee
respectively.

The model (\ref{eq:1}) has a number of interesting
applications to the phase transitions in three-dimensional
simple and complicated systems.
Indeed, when $N=1$ Hamiltonian (\ref{eq:1}) describes
the critical phenomena in pure spin system (Ising model), while
for $N=2$ it corresponds to the anisotropic $XY$ model describing 
structural phase transitions in ferroelectrics as ordering the
two-component alloys \cite{WF72,BLgZ74}.
The magnetic and structural phase transitions in a cubic crystal
are governed by model (\ref{eq:1}) as $N=3$. In the replica limit
$N \to 0$ Hamiltonian (\ref{eq:1}) is known to determine
the critical
properties of weakly disordered quenched systems undergoing second-order
phase transitions \cite{GrLutAh} with a specific set of critical exponents
\cite{HLKh}. Finally, the case $N \to \infty$ corresponds to the
Ising model with equilibrium magnetic impurities \cite{Ah73l}.
In this limit the Ising critical exponent of specific heat $\al$ changes
its sign and takes the Fisher renormalization \cite{F68} together with
$\nu$ and $\gm$: $\al \to - \al/(1 - \al)$, $\nu \to \nu/(1 - \al)$,
$\gm \to \gm/(1 - \al)$.

To calculate the $\bt$-functions normalizing conditions must be imposed
on renormalized one-particle irreducible inverse Green functions
$$\Gm^{(N)}_R (p; m,u,v; \La;d) =
Z_{\vp}^{N/2} \Gm^{(N)} (p; m_0,u_0,v_0; \La;d)$$
given by corresponding Feynman diagrams, $\La$ is the ultraviolet 
momentum cut-off.
Within the massive field-theoretical RG scheme \cite{Par80} at zero
external momenta and at the limit $\La \to \infty$
they are normalized in a conventional way \cite{BLgZ76}:
\beq
\ba{cccc}
\Gm_R^{(2)}(p,-p;m, u, v;d)\Big\arrowvert_{p = 0} &=& m^2 \ , &
\nn \\
{\partial \over {\partial p^2}} \Gm_R^{(2)} (p,-p; m, u, v;d)
\Big\arrowvert_{p = 0} &=& 1 \ , &
\nn \\
\Gm_{u \>R}^{(4)} (\{p_i\}; m, u, v;d) \Big\arrowvert_{\{p_i\} = 0}
&=& m^{4-d} u \ , &
\label{eq:3} \\
\Gm_{v \>R}^{(4)} (\{p_i\}; m, u, v;d) \Big\arrowvert_{\{p_i\} = 0}
&=& m^{4-d} v \ , &
\nn
\ea
\eeq
where $m$, $u$ and $v$ are the renormalized mass and dimensionless
coupling constants. The vertices $\Gm_u^{(4)}$, $\Gm_v^{(4)}$ are
connected with the vertex function without external lines normalized
at zero external momenta
$$
 \Gm_{\>\>ijkl}^{(4)} \> (0) =
 \Gm_u^{(4)} \cdot G^{\> 1}_{\> ijkl}+
 \Gm_v^{(4)} \cdot G^{\> 2}_{\> ijkl}.
$$
From equations (\ref{eq:3}) the expansions for the renormalization constants
$Z_{\vp}$, $Z_u$ and $Z_v$ may be obtained
\beq
\ba{cccc}
Z^{-1}_{\vp} &=& {\partial \over {\partial p^2}}
\Gm^{(2)} (p,-p; m_0, u_0, v_0) \ , &
\nn \\
Z^{-1}_u &=& \frac{1}{u_0} \Gm^{(4)}_u (0; m_0, u_0, v_0) \ , &
\label{eq:4} \\
Z^{-1}_v &=& \frac{1}{v_0} \Gm^{(4)}_v (0; m_0, u_0, v_0) \ . &
\nn \\
\ea
\eeq
These constants relate the "bare" mass $m_0$ and coupling constants
$u_0$ and $v_0$ of the initial Hamiltonian (\ref{eq:1}) to the corresponding
physical parameters
$$
   m_0^2 + \dt m_0^2 = m^2 , \quad m = Z_{\vp}
   \Gm^{(2)} (p,-p; m_0, u_0, v_0) ,
$$
\beq
   u_0 = m^{4-d} \frac{Z_u}{Z^2_{\vp}} u, \quad
   v_0 = m^{4-d} \frac{Z_v}{Z^2_{\vp}} v.
\label{eq:5}
\eeq
With relations (\ref{eq:5}) taken  into account, the $\bt$-functions
can be calculated via the formulae
\beq
\ba{lcr}
\frac{\partial \ln u_0}{\partial u} \bt_u +
\frac{\partial \ln u_0}{\partial v} \bt_v &=& (d-4),
\nn \\
\frac{\partial \ln v_0}{\partial u} \bt_u +
\frac{\partial \ln v_0}{\partial v} \bt_v &=& (d-4),
\label{eq:6}
\ea
\eeq
where $\bt_g \equiv \frac{\partial g}{\partial |\ln m|}$,
$g=\{u,v\}$.

For each Feynman graph contributing to the RG functions the corresponding
contractions are computed by the algorithm developed in Ref. \cite{MV98-1}.
The combinatorial factors as well as the integral values are known from
Ref. \cite{NMB77}. After some work we obtain the four-loop expansions for
the $\bt$-functions in three dimensions:
\be
 \nn
\bt _u &=& u~ \biggl\{1 - u - {\frac 6{{N+8}}}~ v +{\frac 1{{(N+8)^2}}}
 \biggl[3~ \bigl( 2.024691~ N + 9.382716 \bigr)~ u^2 \\
 \nn
 &+& 44.444444~ u~ v  + 10.222222~ v^2 \biggr]
  - {\frac 1{{(N+8)^3}}} \biggl[3~ \bigl(0.449648~ N^2 \\
 \nn
 &+& 18.313459~ N + 66.546806 \bigr)~ u^3
  + 3~ \bigl(6.646878~ N + 164.613849 \bigr)~ u^2~ v \\
 \nn
 &+& 3~ \bigl(0.621889~ N + 100.955929 \bigr)~ u~ v^2
  + 65.937285~ v^3 \biggr] \\
 \nn
 &+& {\frac 1{{(N+8)^4}}}
 \biggl[- \bigl(0.155646~ N^3 - 35.820204~ N^2 - 602.521231~ N \\
 \nn
 &-& 1832.206732 \bigr)~ u^4 - 3~ \bigl(1.352882~ N^2 - 182.073890~ N \\
 \nn
 &-& 2064.170701 \bigr)~ u^3~ v + 3~ \bigl(27.250336~ N
  + 2110.408809 \bigr)~ u^2~ v^2 \\
 &+& 9~ \bigl(1.291017~ N + 308.599361 \bigr)~ u~ v^3
  + 495.005747~ v^4 \biggr] \biggr\} \ ,
\label{eq:BU}
\ee
\be
 \nn
\bt _v &=& v~ \biggl\{1 - {\frac 1{{N+8}}}\bigr(12~ u + 9~ v \bigr)
  +{\frac 1{{(N+8)^2}}}
 \biggl[\bigl( 3.407407~ N + 54.814815 \bigr)~ u^2 \quad \\
 \nn
 &+& 92.444444~ u~ v + 34.222222~ v^2 \biggr]
  - {\frac 1{{(N+8)^3}}} \biggl[- \bigl(1.251107~ N^2 \\
 \nn
 &-& 41.853902~ N - 469.333970 \bigr)~ u^3 + 9~ \bigl(0.248784~ N
  + 136.511768 \bigr)~ u^2~ v \\
 \nn
 &+& 957.781662~ u~ v^2 + 255.929737~ v^3 \biggr]
  + {\frac 1{{(N+8)^4}}} \biggl[\bigl(0.574653~ N^3 \\
 \nn
 &-& 0.267107~ N^2 + 584.287672~ N + 5032.692260 \bigr)~ u^4
  + 3~ \bigl(0.057375~ N^2 \\
 \nn
 &+& 107.641680~ N + 5989.283536 \bigr)~ u^3~ v
  + 3~ \bigl(7321.464604 \\
 &-& 16.494003~ N \bigr)~ u^2~ v^2 + 11856.956858~ u~ v^3
  + 2470.392521~ v^4 \biggr] \biggr\} \ .  \label{eq:BV}
\ee

These equations are known to have four solutions corresponding to 
the trivial Gaussian, the Ising, the isotropic (Heisenberg) and 
the cubic fixed points \cite{Ah73}. The most intriguing of them
are the isotropic and cubic ones. 
The one-loop approximation analysis of the eigenvalue exponents for 
these points yields the upper boundary value for the critical 
dimensionality $N_c$, $N_c = 4$.

To determine the lower boundary of $N_c$ one should attract the specific
symmetry property of model (\ref{eq:1}), when $N=2$ \cite{MS87}.
Namely, the transformation of the field components
\be
\vp_1 \rightarrow {1\over {\sqrt{2}}}~(\vp_1
+ \vp_2) \ , \qquad
\vp_2 \rightarrow {1\over {\sqrt{2}}}~(\vp_1 -
\vp_2) \label{eq:9}
\ee
combined with substitution of the quartic couplings
\be
u \rightarrow u + {3 \over 2}~v \ , \quad v \rightarrow - v
\label{eq:10}
\ee
does not change the structure of the initial Hamiltonian itself.
As a result, the $\bt$-functions (\ref{eq:BU}), (\ref{eq:BV})
should obey certain symmetry relations \cite{Kor76}:
\be
\bt_u \biggl( u + {3 \over 2}~ v, - v \biggr) &=&
\bt_u ( u, v) + {3 \over 2}~ \bt_v ( u, v) \ ,
\nn \\
\bt_v \biggl( u + {3\over 2}~ v, - v \biggr) &=&
- \bt_v ( u, v) \ , \quad \label{eq:11}
\ee
but their form remain unchanged. However, for $N=2$ transformations
(\ref{eq:9}) and (\ref{eq:10}) result in the relocation
of the coupling constants values so that the cubic and Ising fixed
points are transformed into each another at the 3D RG flow diagram.
Since the exact RG equations always have the Ising fixed point, which
inevitably is the saddle-knot one, these equations
should also have the cubic fixed point, which will be unstable. In this
situation, the isotropic fixed point, again always existing in the exact
RG equations, should be the stable knot only. Therefore, we conclude that
the lower boundary of $N_c$ is not less than two.
The real value of $N_c$ can be obtained only on the basis of analysis
of the RG flow diagram structure, provided the $\bt$-functions of the model
are calculated in sufficiently high-order RG approximations and then
processed by appropriate resummation techniques.

\section{Resummation, fixed points and stability}
\label{sec:3}

It is well known that the field-theoretical RG series are divergent.
The character of their large-order behaviour is well established only
for simple $O(N)$-symmetric models \cite{Lip77,BLgZ77,BrPr78}.
The coefficients of the series at large $k$ were shown to behave as
$c (-a)^k k! k^b$, where the asymptotic parameters $a$, $b$ and $c$
are assumed to be calculated for each RG function. Knowledge of the exact
values of the asymptotic parameters in combination with the most powerful
resummation procedure of the Borel transformation with a conformal
mapping, first proposed in Ref. \cite{LgZ77}, made it possible
to develop the accomplished quantitative theory of critical behaviour of
simple systems \cite{GdZ98,Z}.

At the same time, the asymptotic nature of RG functions of anisotropic
models is still unknown. Calculating the large-order asymptotic behaviour
for the series in such models is a very difficult problem.
That is why, lacking any information about the large-order behaviour
either the simple
Pad$\acute{\rm e}$-Borel or Chisholm-Borel resummation procedures are used.
The latter technique, however, possesses at least two inherent drawbacks.
First, some ambiguity in the calculation of coefficients
of denominators of the Chisholm approximants is unavoidable \cite{Ch73}.
Second, the Chisholm-Borel procedure does not hold the specific symmetry
properties of a model. At the same time, exploiting the Borel
transformation in combination with the Pad$\acute{\rm e}$ or Chisholm
approximants shows that the results of calculation are very sensitive to
the choice of the type of approximants.
This may lead to estimates which do not provide
reliable predictions even in the higher-loop RG approximations \cite{SV99}.
Besides, in the framework of both schemes it is very difficult to determine
any error bounds for the evaluated quantities.

In this paper we attempt to overcome the outlined difficulties by applying 
the PBL resummation method, generalized for the two coupling constant case,
to processing the RG expansions (\ref{eq:BU}) and (\ref{eq:BV}). This method, 
first introduced by Baker {\it et al} in \cite{BNM78}, turned out to be 
highly efficient when used
to study the critical behaviour of the simple $O(N)$-symmetric models in 3D.
The critical exponent estimates obtained within the framework of this
technique are regarded nowadays as the most accurate values, as those of
\cite{LgZ77,GdZ98,LgZ80}. We motivate our choice of the PBL resummation
method with the following reasons.
\begin{itemize}
\item 3D RG expansions for the $\bt$-functions of the cubic model
      alternate in signs. Therefore, using PBL resummation technique 
      is quite natural.
\item It can be expected that for complex models with more than one
      coupling constant, the asymptotics of the RG series at large orders
      will include a factor $k! k^b$. The PBL resummation method removes
      divergences of this type.
\item The PBL resummation method allows one to determine the error bounds
      for the physical quantities to be calculated, in a natural way.
\end{itemize}

\nid
The generalized PBL resummation procedure consists of the following.
Let a physical quantity $F(u,v)$ be represented by a double series
\be
F(u, v) = \sum_{i,j} f_{ij} u^i v^j \ ,
\label{eq:12}
\ee
where coefficients $f_{ij} \sim (i+j)! (i+j)^b$ at large orders
$(i,j \to \infty)$, the additional  parameter $b$ being an arbitrary
non-negative number to be defined below. Associated with the initial
series (\ref{eq:12}) is the function
\be
{\cal F}(u, v; b) = \int\limits_0^\infty
e^{-t} t^b B(ut, vt) dt \ \
\label{eq:13}
\ee
The Borel-Leroy transform $B(x,y)$ is the analytical
continuation of its Taylor series
\be
B(x, y) = \sum_{ij} {f_{ij} \over {\Gm(i+j+b+1)}} x^i y^j
\label{eq:14}
\ee
which is
absolutely convergent in a circle of non-zero radius. In order to
calculate the integral in (\ref{eq:13}) one should continue analytically
$B(x,y)$ for $0 \le x < \infty$ and $0 \le y < \infty$.
To this end, the rational Pad$\acute{\rm e}$ approximants [L/M] $(x,y)$
are used. The Pad$\acute{\rm e}$ approximant method is determined in
a conventional way \cite{BGm}. Let us consider a "resolvent" series
\be
\tilde B (x, y, \la) = \sum_{k = 0}^{\infty} \la^k
\sum_{l = 0}^k {f_{l, k-l} x^l y^{k-l}
\over {\Gm(k+b+1)}} = \sum_{k = 0}^{\infty} A_k \la^k \ ,
\qquad \qquad
\label{eq:15}
\ee
where coefficients $A_k$ are uniform polynomials of $k$th order in $u$
and $v$. The sum of the series is then approximated by
\be
B (x, y) = [ L / M ] \Big \arrowvert_{\la = 1} \ .
\label{eq:16}
\ee
The Pad$\acute{\rm e}$ approximants [L/M] in $\la$ are given by 
\be
[L/M] = {P_L (\la) \over {Q_M (\la)}} \ , \label{eq:16a}
\ee
where $P_L (\la)$ and $Q_M (\la)$ are polynomials of degrees $L$
and $M$, respectively, with coefficients depending on $x$ and $y$ which
should be determined from the conditions
\be
&Q_M&(\la) \tilde B (x, y; \la) - P_L (\la) =
O (\la^{L + M + 1}) \ , \nn \\
&Q_M&(0) = 1 \ .
\label{eq:17}
\ee
Replacing variables $x = u t$ and $y = v t$ in the Pad$\acute{\rm e}$
approximants and then evaluating the Borel-Leroy integral
\be
{\cal F} (u, v; b) = \int\limits_0^{\infty} e^{-t} t^b [L/M]
\Big \arrowvert_{\la = 1} dt \ 
\label{eq:18}
\ee
we obtain the approximate expressions for RG functions.

Among Pad$\acute{\rm e}$ approximants the diagonal ($L = M$) or
near-diagonal ones were proved to exhibit the best approximating
properties \cite{BGm}.
However, as the degree of the denominator $M$ increases, the number of
possible poles of the approximant increases too. If some of the
poles belong to the positive real semiaxis, the
corresponding approximant should be rejected. Due to this the choice of
"working" approximants, which might be used for analytical continuation
of the Borel-Leroy image onto the complex cut plane, is largely limited.
On the other hand, varying the free parameter $b$ in the Borel-Leroy
transformation (\ref{eq:13}) allows one to optimize the resummation
procedure under the condition that the fastest convergence of the iteration
process is achieved. So, taking into account the above-mentioned remarks,
in order to find the locations of the fixed points 
we adopt the following scheme. For
the fixed $N$, the $\bt$-functions are resummed by virtue of transformation
(\ref{eq:13}) in the highest-loop orders by shifting the transformation
parameter $b$. For an analytical continuation of the
Borel-Leroy transforms $B_u (u,v)$, $B_v (u,v)$ over the cut plain the most
appropriate Pad$\acute{\rm e}$ approximants [2/1], [3/1] and [2/2]
are used. The locations of the fixed points are then determined  
for each $b$ from the solution
of the set of equations: $\bt^{res}_u (u_c, v_c) = 0$,
$\bt^{res}_v (u_c, v_c) = 0$. The "true" locations are obtained
by averaging over the values given by the approximants under the optimal
value of the parameter $b$, at which the quantity
$|1 - {\cal F}_L (u,v;b)/{\cal F}_{L-1} (u,v;b)|$ reaches its local minima.
The quantity ${\cal F}_L (u,v;b)$ is evaluated for the $L$-partial sum 
of the series in equation (\ref{eq:18}), where $L$ denotes
the step of truncation of the series.

The results of the computation of the cubic fixed point locations depending 
on the parameter $b$ are presented for the physically important case
$N = 3$ in figure 1. Three curves correspond to the three Pad$\acute{\rm e}$
approximants. The parameter $b$ shifts from $0$ to $3$. As can be seen from
the figure the optimal value of $b$ is zero. At this point the
numerical values of the cubic fixed point locations given by different
approximants are the closest to each other.
The result of computing the cubic fixed point locations for $N=3$ are
also presented in table 1. In the first three columns of the table the fixed
point locations values found by means of the Pad$\acute{\rm e}$ approximants
[2/1], [3/1], and [2/2] at $b=0$ are given. Averaging the results of 
processing over all of the approximants under the optimal
value of $b$ gives the estimates presented in the fourth column of the table.
We adopt these numbers as the final estimates for the cubic fixed point
locations found within the four-loop approximation. As the degree of accuracy 
for these approximate values we take the maximum deviations
of the average values of the fixed point locations from those given by the
approximants at $b=0$.

One can observe, looking at figure 1,
that the values of the cubic fixed point locations given by the symmetric
approximant [2/2] depend weakly on the shift parameter $b$. Averaging over 
all the values given by this approximant within the interval [0,3] results in
the cubic fixed point locations estimates presented in the fifth column
of table 1.
The coordinates of the cubic fixed point found earlier on the basis of the
three- and four-loop approximations, using the Chisholm-Borel resummation 
method, are presented in the table, for comparison. 
These numbers include the normalizing multiplier
$11 \over 9$ needed to compare our $\bt$-functions with those obtained
in \cite{Sp89,MSSh89}.

To verify the correctness of the chosen approach let us apply the 
above considered scheme to estimate the fixed point locations of the
$O(3)$-symmetric model for which the numerical results are well known.
The six-loop 3D RG expansion for the $\bt$-function of this model was
reported in \cite{LgZ77,BNM78}. The PBL resummation of that
series using eight types of Pad$\acute{\rm e}$ approximants
[2/1], [3/1], [2/2], [4/1], [3/2], [5/1], [4/2] and [3/3] yields,
after solving the equation $\bt^{res} (g_c) = 0$, the picture displayed
in figure 2. It is seen that the values of the isotropic fixed point location
calculated in the highest RG orders with the help of the approximants [3/3],
[4/2] and [3/2] are very weakly dependent on the parameter $b$ varied
within the interval $0 \le b \le 15$.
The curves corresponding to these approximants are intersected at the
point $b=4.5$. Therefore $b=4.5$ is the optimal value of the transformation
parameter for which the fastest convergence of the iteration procedure
is ensured. For $b=4.5$ the central value estimate
of the isotropic fixed point is $g_c = 1.392$. The maximum deviation of
the central value from the values given by some of the approximants [3/3],
[4/2] and [3/2] at the point $b=10$ is adopted approximately as an apparent
accuracy of the calculation, $\Dt = 0.0013$. Such a small error can be
explained by the small dispersion of the curves within
the range $5 \le b \le 10$. So, the estimate $g_c = 1.3920 \pm 0.0013$
is in excellent agreement with those found more then 20 years ago in
\cite{LgZ77,BNM78} as well as with recent results of \cite{GdZ98}.

Within the framework of the four-loop approximation there are only three
appropriate Pad$\acute{\rm e}$ approximants.
Averaging the results of computing the isotropic fixed point location
given by the approximants [2/1], [3/1] and [2/2] under the optimal value
of the transformation parameter results in the estimate
$g_c = 1.3925 \pm 0.0070$. The error was determined again through the maximum
deviation of the central value from those given by each of the approximants
at $b=0$. It is seen that the four-loop estimate of the coordinate of the
isotropic fixed point is in a good accordance with the best ones followed
from the six-loop consideration.

Note, however, that the coordinate of the $O(3)$-symmetric fixed point
calculated within the five-loop approximation does not approach the "exact"
value. Namely, the PBL resummation procedure leads to the estimate
$g_c = 1.3947 \pm 0.0040$. Although the error of the calculation became
visibly smaller, the central value of the fixed point location stepped
aside from the four- and six-loop ones.

Thus, the fulfilled numerical
analysis shows that the isotropic fixed point location estimate obtained
in the four-loop level occurs close to the six-loop value.
One can expect, therefore, that in the case of the cubic model the
fixed point locations $u_c=1.3428 \pm 0.0200$, $v_c=0.0815 \pm 0.0300$
(fourth column of table 1) will not be strongly distinguish from the "exact",
say, the six-loop, values. The coordinates of the cubic fixed point for 
some $N$ are presented in table 2.
Our calculations show that for $N=3$ the coordinates of the cubic fixed point
practically do not differ from those of the Heisenberg one. However, with
increasing $N$ the cubic fixed point runs away from the isotropic
point moving towards the Ising one. In the large $N$ limit these two fixed
points become close to each another so much that the influence of the
$O(N)$-symmetric invariant on the critical thermodynamics of the cubic model
vanishes. This can be easily seen by applying the
$\frac{1}{N}$ consideration to the one-loop solutions of the RG equations
of model (\ref{eq:1}). Indeed, rescaling the coupling constants
$u \rightarrow u/N$, $v \rightarrow v/N$ in the initial Hamiltonian and
taking then the limit $N \to \infty$ one can see that the cubic fixed point 
approaches the Ising one asymptotically.
So, the cubic model turns out to be split into $N$ non-interacting Ising
models, the critical behaviour of each of them will be determined by a set of
the critical exponents renormalized according to Fisher \cite{F68}.

Another way to determine the fixed point locations in fixed $D$
is to construct the RG flows diagram of the model. If, at the flows
diagram, there exists a fixed point of stable knot type, the trajectories
originated from some point within the range of stability of the initial
Hamiltonian would flow towards the knot. The region at the flow diagram
where the trajectories are intersected provides the coordinates of
the stable fixed point. Investigating the 3D RG flow diagram of 
model (\ref{eq:1}) in the four-loop approximation we arrive at the conclusion
that the cubic rather than the isotropic fixed point is absolutely stable for
all $N \ge 3$.

At the same time, the reliable prediction about the stability of the
cubic fixed point for $N \ge 3$ can be made on the basis of calculating
the eigenvalue exponents $\la$'s of the stability matrix
\lmx{M_{ij}}
    \frac{\partial\bt_u}{\partial u} &
    \frac{\partial\bt_u}{\partial v} \\
    \frac{\partial\bt_v}{\partial u} &
    \frac{\partial\bt_v}{\partial v}
\rmx
taken at $u=u_c$ and $v=v_c$. If the real parts of both eigenvalues are
negative, the fixed point is the stable knot in the $(u,v)$ plane.
If $\la_1$, $\la_2$ have opposite signs, the point is of
the "saddle-knot" type.

To calculate the stability matrix eigenvalues of the cubic and isotropic
fixed points we have chosen the following strategy. First, the derivatives of
the $\bt$-functions (\ref{eq:BU}) and (\ref{eq:BV}) are calculated, 
and the new 
RG expansions resummed by means of the PBL technique are substituted into
the matrix $M_{ij}$. The eigenvalue exponents of the matrix of derivatives
$M_{ij}$ obtained in such a way are then evaluated under the optimal value
of the transformation parameter $b$. In figure 3 we present our
numerical results for the series $-\frac{\partial\bt_u}{\partial u}$ and
$-\frac{\partial\bt_v}{\partial v}$ for the physically interesting
case $N=3$. The curves correspond to the three types of the
Pad$\acute{\rm e}$ approximants used within the four-loop approximation.
The crossing of the curves gives the optimal value of $b$ at which
we find $\frac{\partial\bt_u}{\partial u}|_{opt} = -0.7536$ and
$\frac{\partial\bt_v}{\partial v}|_{opt} = -0.0331$. Because the series
$-\frac{\partial\bt_u}{\partial v}$ and $-\frac{\partial\bt_v}{\partial u}$
turn out to be shorter by one order in comparison with
$-\frac{\partial\bt_u}{\partial u}$ and $-\frac{\partial\bt_v}{\partial v}$,
their resumming performed with the help of the approximant [2/1] only yields
the monotonic dependence of the result of processing on the parameter $b$.
In this unfavorable situation, we take into account an additional
Pad$\acute{\rm e}$ approximant [1/1] to optimize the iteration procedure.
The results are plotted in figure 4. For the optimal values of
$b$ we obtain $\frac{\partial\bt_u}{\partial v}|_{opt} = -0.4566$
and $\frac{\partial\bt_v}{\partial u}|_{opt} = -0.0409$. Straightforward
calculation of the eigenvalues of the stability matrix $M_{ij}$ gives for
the cubic fixed point the numbers provided in table 3.
The eigenvalues of the isotropic fixed point as well as the analogous
numerical estimates obtained recently in \cite{KTSf-b} on the basis of
using the five-loop $\ve$-expansions are presented for comparison therein.
These estimates show that the cubic fixed point is absolutely stable in 3D
for $N=3$ while the isotropic fixed point appears to be stable on the
$u$-axis only. Our numerical results agree well with those obtained in 
\cite{KTSf-b}. Unfortunately, at present we cannot indicate realistic
error bounds in our calculation of the stability matrix eigenvalues.
Nevertheless, a crude estimate can be done. In fact, if the model
(\ref{eq:1}) is almost identical to some marginal system for which $N_c=3$
in 3D, the stability index $\la_2$ both for the isotropic and for the cubic
fixed points should be equal to zero at $N=3$ and, consequently, the
points should coincide. Therefore, as can be seen from the numbers given 
in table 3, the four-loop approximation predicts eigenvalues with error 
about 0.01. Of course, dealing with the theory without a small parameter 
and the short perturbative expansions one would refer to such a level of 
accuracy as satisfactory. However numerically small
errors may lead, sometimes, to qualitatively incorrect results \cite{SV99}.

Note, that although the recent high precision MC simulation
using finite size scaling techniques \cite{CH98} predicts the stability
of the isotropic rather than the cubic fixed point, the absolute value of
the stability eigenvalue $|\la_2|$ obtained for the isotropic point turned
out to be very small. This is in accordance with our estimate.

Let us now calculate the critical dimensionality $N_c$ of the order parameter
field. The critical dimensionality is defined as a value of $N$ at
which the cubic fixed point coincides with the isotropic one. Equivalently,
for $N=N_c$ the second eigenvalue of the stability matrix $M_{ij}$ vanishes,
$\la_2 = 0$. Studying carefully the 3D RG flow diagram of model
(\ref{eq:1}) depending on the order of approximation for different 
Pad$\acute{\rm e}$ approximants we
arrive at the conclusion that $N_c=2.910 \pm 0.035$ and $N_c=2.890 \pm 0.020$
within the three- and four-loop approximations, respectively. The accuracy of
the calculation of $N_c$ was determined through the evaluation of the stability
matrix eigenvalues for different $N$ from the interval of errors mentioned
above.
That value of $N=N_c$, above or below its central number, at which the
second eigenvalue $\la_2$ was becoming non-zero, was taking for the upper or
lower boundary of $N_c$, respectively.

It is worth comparing the four-loop estimate of $N_c$ just found with that
which can be obtained within the $\ve$-expansion method. The five-loop
$\ve$-expansion for $N_c$ has been calculated in \cite{KS95}.
The series turned out to be alternating in signs that allows one to resum
it by means of the PBL technique. To this end, we use again
the most appropriate
Pad$\acute{\rm e}$ approximants [2/1], [3/1] and [2/2] for analytical
continuation of the Borel-Leroy transform for all $0 \leq \ve t \le \infty$.
Dependence of the results of processing of the critical
dimensionality $N_c$ on the transformation parameter $b$ is depicted in
figure 5. The curves corresponding to the approximants are crossed at
the point $b \sim 1$. The appropriate value of the critical dimensionality
is $N_c=2.894 \pm 0.040$. As an error of the calculation it is natural
to assume the maximum scattering of numerical values given by the 
approximants at $b=0$ from that
obtained at the crossing point of the curves. This estimate of $N_c$
is in excellent accordance with the above found within the 3D RG approach.
It also agrees well with the estimates obtained earlier on the basis of the
different resummation technique \cite{Sp89,MSSh89}.
So, both schemes, the RG technique directly in 3D and the $\ve$-expansion
method, result in the same estimate of the critical dimensionality $N_c=2.89$,
thus implying that the cubic fixed point is stable in three dimensions for
$N \ge 3$. This means that the critical behaviour of the magnetic phase
transitions in crystals with cubic anisotropy should belong to the cubic
rather than the isotropic universality class with a certain set of critical
exponents. However, due to the obvious marginality of the model
($N_c \sim 3$) and closeness of both (isotropic and cubic) fixed points
on the 3D RG flow diagram for $N=3$, the critical exponents values of
the cubic point will be practically the same of the isotropic one. That is
why the calculation of the critical exponents in cubic magnets with $N=3$
seems to be of academic interest only.

\section*{Conclusion}

To summarize,
the complete analysis of the global structure of RG flows of a
model with two quartic coupling constants associated with isotropic and
cubic interactions describing magnetic and
structural phase transitions in a good number of real substances has been
carried out within the massive field theory directly in three dimensions.
Perturbative expansions for the $\bt$-functions were deduced for generic $N$
up to four-loop order. The fixed points locations were found for $N\ge3$
by applying the generalized Pad$\acute{\rm e}$-Borel-Leroy resummation
technique.
On the basis of comparative numerical analysis with the $O(N)$-symmetric
models, the fixed point locations for which have been solidly established
\cite{GdZ98,Z}, we have made the assumption that the four-loop estimates
of the cubic fixed point locations should not differ strongly from
the "exact" values within the error bounds.

The analysis of the eigenvalue exponents of the isotropic and cubic fixed
points fulfilled for the physically significant case $N=3$ has shown that
the cubic rather than the isotropic fixed point is absolutely stable in 3D.
The eigenvalues estimates (see table 3) were found to agree well with those
calculated on the basis of exploiting the five-loop $\ve$-expansions
combined with a careful resummation procedure \cite{KTSf-b}. Our results 
agree numerically with the recent high precision MC estimates \cite{CH98}, 
in spite of the latter predicting the stability of the isotropic rather than 
the cubic fixed point.

The critical dimensionality $N_c$ of the order parameter,
at which the topology of the flow diagram changes, has been estimated by
the two different methods:
(a) by resumming the four-loop RG expansions for the $\bt$-functions
in 3D and (b) by resumming the five-loop $\ve$-expansion for $N_c$ at
$\ve=1$. The numerical estimates $N_c=2.89 \pm 0.02$ and 
$N_c=2.894 \pm 0.040$
obtained are in a good agreement with the earlier results \cite{Sp89,MSSh89}
and confirm the conclusion about the stability of the cubic fixed point for
$N\ge3$. Consequently, the magnetic and structural phase transitions in
three-dimensional anisotropic crystals with cubic symmetry are of second order
and their critical thermodynamics should be governed by the cubic fixed point
with a certain set of critical exponents.
Unfortunately, the cubic universality class is not easily distinguished
experimentally from the isotropic one, due to the obvious marginality
of the problem, $N_c \sim 3$.

\nid

\vspace{1.5cm}

\newpage
\begin{center}
{\large \bf Figure Captions}
\end{center}

\vspace{0.5cm}

\nid
Fig. 1. Curves demonstrating the dependence of the results of calculating
the cubic fixed point locations on the transformation parameter 
$b$ for $N = 3$.
The upper curve ($\Diamond$) corresponds to the [2/1] approximant, while the
middle ($\triangle$) and lower ($\Box$) curves correspond to the [2/2] and
[3/1] approximants, respectively.

\vspace{0.5cm}

\nid
Fig. 2. The results of computation of the O(3)-symmetric fixed point
locations from the three- to the six-loop approximations obtained on
the basis of the PBL resummation method with eight types of the
approximants: [2/1] - $\Diamond$,
[3/1] - $\Box$, [2/2] - $\triangle$, [4/1] - full $\Diamond$,
[3/2] - full $\triangle$, [5/1] - full $\circ$, [4/2] - $\circ$,
[3/3] - $\times$.

\vspace{0.5cm}

\nid
Fig. 3. Graphs of dependence of the results of processing of the series
a) $-\frac{\partial\bt_u}{\partial u}$,
b) $-\frac{\partial\bt_v}{\partial v}$
on the parameter $b$, $N=3$. The curves are given in the 
same notations as in the previous figures.

\vspace{0.5cm}

\nid
Fig. 4. Graphs of dependence of the results of processing of the series
a) $-\frac{\partial\bt_u}{\partial v}$,
b) $-\frac{\partial\bt_v}{\partial u}$
on the parameter $b$, $N=3$. For the curves corresponding to the 
approximants [1/1] and [2/1] the notation $\Diamond$ and $\Box$ are used, 
respectively.

\vspace{0.5cm}

\nid
Fig. 5. Dependence of the results of processing of the $\ve$-series for
the critical dimensionality $N_c$ on the transformation parameter $b$.

\begin{table}[1]
\caption{Coordinates of the cubic fixed point of RG equations
 for $N=3$ found under the optimal value of the transformation
 parameter $b=0$.}
\vspace{0.5cm}
\hspace{0.8cm}
\begin{tabular}{|c|c|c|c|c|c|}
\hline
& [2/1] & [3/1] & [2/2] & Average value & Average over [2/2] \\
\hline
\hline
$u_c$ & 1.3536 & 1.3338 & 1.3410 & 1.3428 $\pm$ 0.0200     
      & 1.3425 \\
      &        &        &        & 1.3480$^a$, 1.3357$^b$ 
      & \\
\hline
$v_c$ & 0.0526 & 0.1026 & 0.0894 & 0.0815 $\pm$ 0.0300 
      & 0.0937 \\
      &        &        &        & 0.0904$^a$, 0.0906$^b$ 
      & \\
\hline
\multicolumn{5}{l}{\footnotesize $^a$ Quoted from
Ref. \cite{Sp89}} \\
\multicolumn{5}{l}{\footnotesize $^b$ Quoted from
Ref. \cite{MSSh89}} \\
\end{tabular}
\label{table 1}
\end{table}
\begin{table}[2]
\caption{Coordinates of the cubic fixed point of RG equations for some
 $N$ found under the optimal value of the transformation parameter $b$
 within the four-loop approximation. The average values of the coordinates
 calculated over the most stable Pad$\acute{\rm e}$ approximant [2/2] 
 are also presented for comparison.} 
\vspace{0.5cm}
\hspace{0.25cm}
\begin{tabular}{|c|c|c|c|c|c|c|c|}
\hline
 N          & 4      & 5      & 6      & 7      & 8
& 9         & 10     \\
\hline
\hline
$u_c$       & 0.9055 & 0.6980 & 0.5807 & 0.5060 & 0.4544
   & 0.4168 & 0.3881 \\
\hline
$v_c$       & 0.8167 & 1.2361 & 1.5386 & 1.7874 & 2.0076
   & 2.2108 & 2.4032 \\
\hline
$u_c$ [2/2] & 0.8981 & 0.6886 & 0.5708 & 0.4962 & 0.4448 
   & 0.4074 & 0.3789 \\
\hline
$v_c$ [2/2] & 0.8380 & 1.2608 & 1.5649 & 1.8148 & 2.0359 
   & 2.2400 & 2.4333 \\
\hline
\end{tabular}
\label{table 2}
\end{table}
\begin{table}[3]
\caption{Four-loop eigenvalue exponents estimates for the cubic (CFP)
 and isotropic (IFP) fixed points found for $N=3$ under the
 optimal value of the transformation parameter $b$.}
\vspace{0.5cm}
\hspace{1.5cm}
\begin{tabular}{|c|c|c|c|c|}
\hline
& CFP   & CFP, Ref. \cite{KTSf-b} & IFP & IFP, Ref. \cite{KTSf-b}  \\
\hline
\hline
$\la_1$ & -0.7786  & -0.7648  & -0.7791 & -0.7640 \\
\hline
$\la_2$ & -0.0081  & -0.0085  & 0.0077  &  0.0089 \\
\hline
\end{tabular}
\label{table 3}
\end{table}


\begin{thebibliography}{99}
\bibitem{remark} It was established that the diagonal or near-diagonal
Pad$\acute{\rm e}$ approximants exhibit the best approximating properties.
See for instance \cite{BGm}
\end{thebibliography}
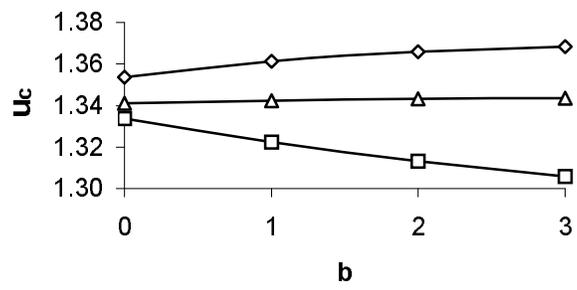
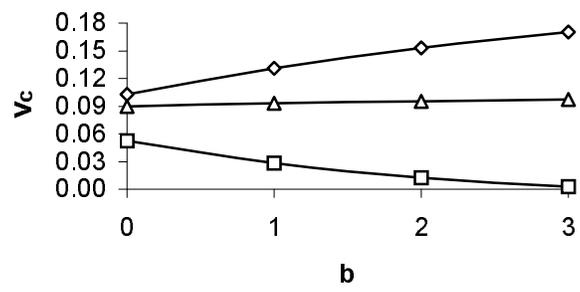
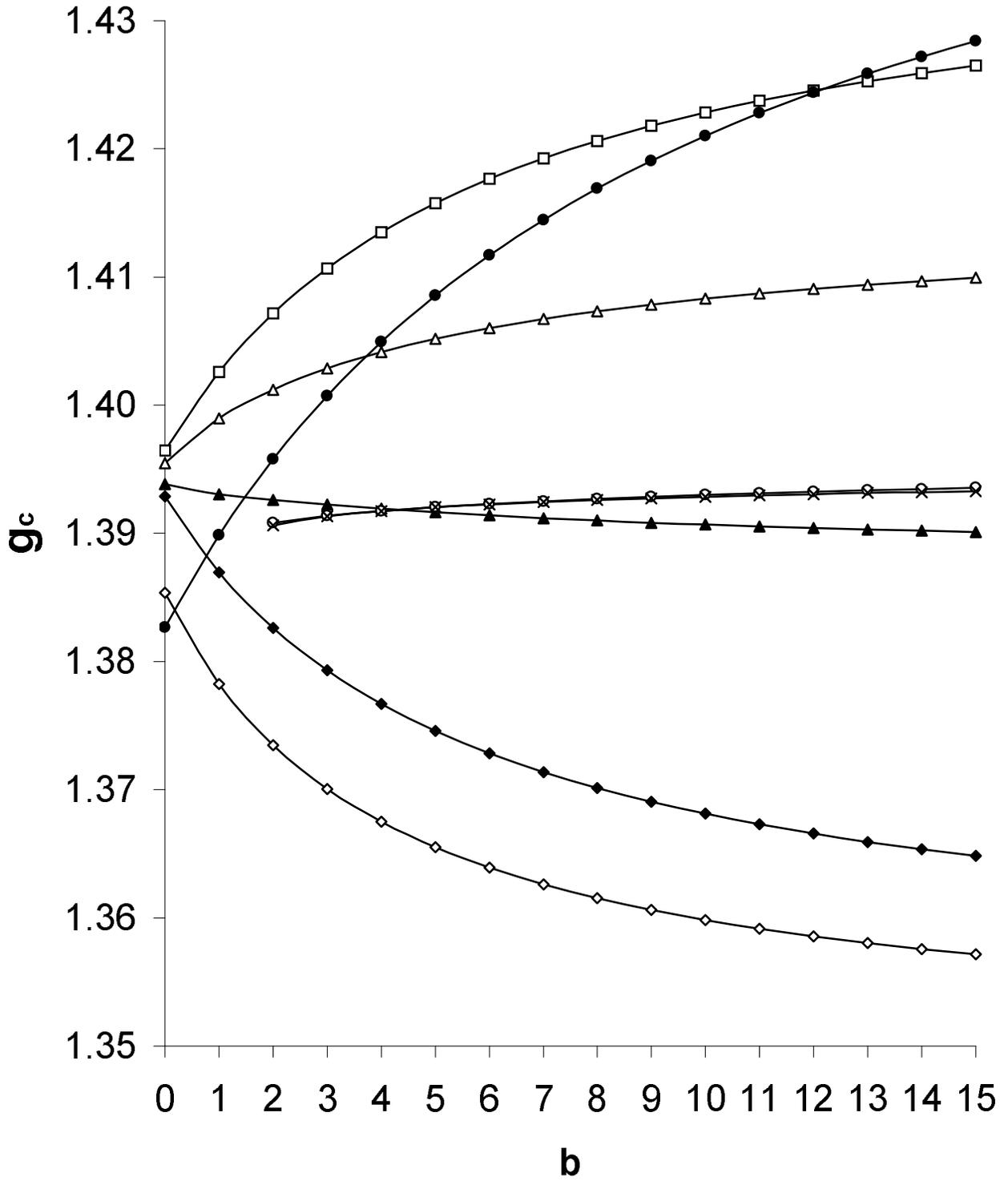
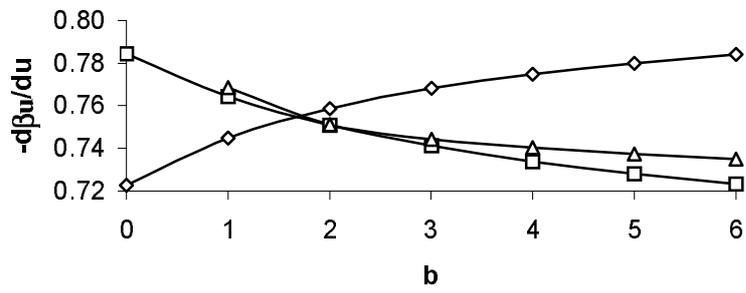
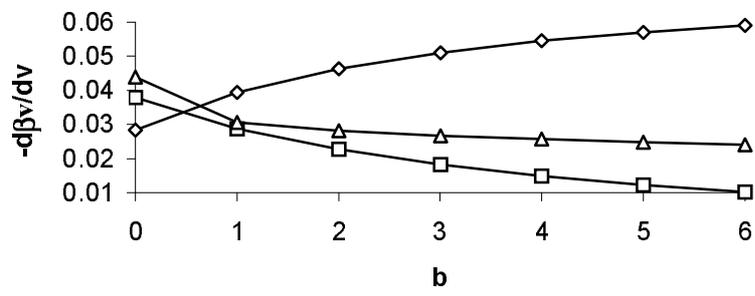
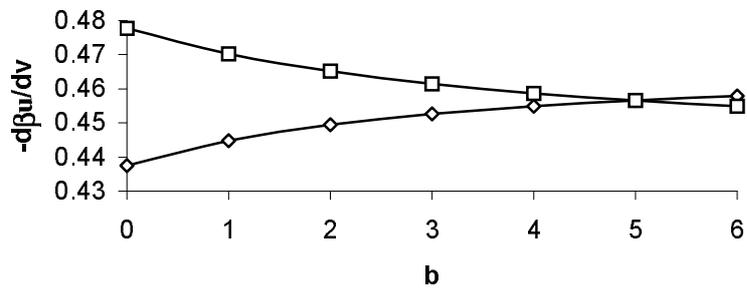
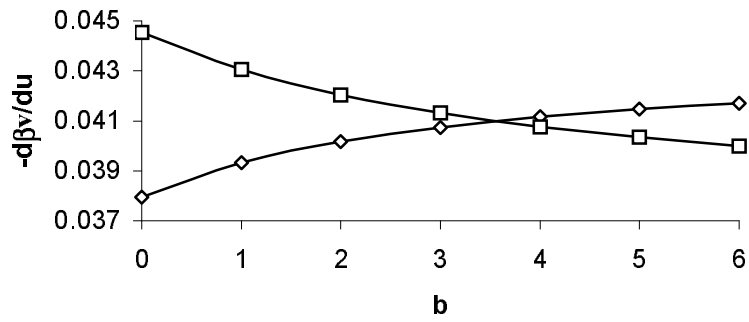
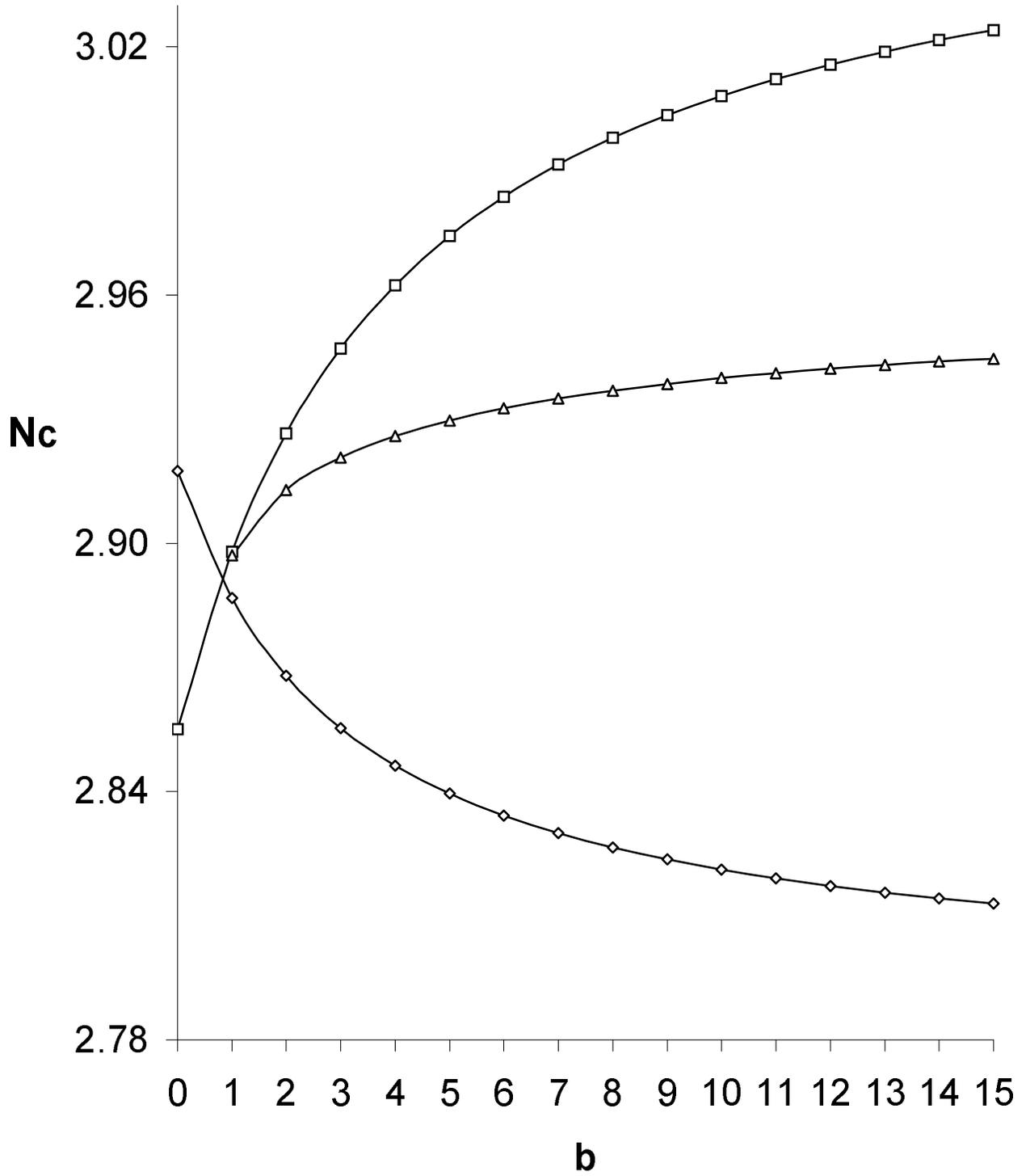
\end{document}